\documentclass[twocolumn,superscriptaddress,pre]{revtex4-1}

\usepackage[latin1]{inputenc}
\usepackage{mathtools}
\usepackage{amsfonts}
\usepackage{amssymb}

\usepackage{subfigure}

\usepackage{microtype}

\usepackage{graphicx, color}

\begin{document}

\title{Zipf and Heaps laws from dependency structures in component systems}

\author{Andrea Mazzolini}
\affiliation{Dipartimento di Fisica and INFN, Universit\`a degli Studi di Torino,
Via Pietro Giuria 1, 10125 Torino, Italy}

\author{Jacopo Grilli}
\affiliation{Santa Fe Institute, 1399 Hyde Park Road, Santa Fe, NM 87501, USA}

\author{Eleonora De Lazzari}
\affiliation{Sorbonne
  Universit\'es, UPMC Univ Paris 06, UMR 7238, Computational and
  Quantitative Biology, 4 Place Jussieu, Paris,
  France} 

\author{Matteo Osella}
\affiliation{Dipartimento di Fisica and INFN, Universit\`a degli Studi di Torino,
Via Pietro Giuria 1, 10125 Torino, Italy}

\author{Marco Cosentino Lagomarsino}
\affiliation{Sorbonne
  Universit\'es, UPMC Univ Paris 06, UMR 7238, Computational and
  Quantitative Biology, 4 Place Jussieu, Paris,
  France} 
  \affiliation{CNRS, UMR 7238, Paris, France}
  \affiliation{IFOM, Milan, Italy}

\author{Marco Gherardi} \email{Corresponding author: marco.gherardi@mi.infn.it}
\affiliation{Sorbonne
  Universit\'es, UPMC Univ Paris 06, UMR 7238, Computational and
  Quantitative Biology, 4 Place Jussieu, Paris,
  France} 
\affiliation{Dipartimento di Fisica, Universit\`a degli Studi di
  Milano, via Celoria 16, 20133 Milano, Italy}

\begin{abstract}
  Complex natural and technological systems can be considered, on a
  coarse-grained level, as assemblies of elementary components: for
  example, genomes as sets of genes, or texts as sets of words.  On
  one hand, the joint occurrence of components emerges from
  architectural and specific constraints in such systems. On the other
  hand, general regularities may unify different systems, such as the
  broadly studied Zipf and Heaps laws, respectively concerning the
  distribution of component frequencies and their number as a function
  of system size.
  Dependency structures (i.e., directed networks encoding the
  dependency relations between the components in a system) were
  proposed recently as a possible organizing principles underlying
  some of the regularities observed.
  However, the consequences of this assumption were explored only in
  binary component systems,
  where solely the presence or absence of components is considered,
  and multiple copies of the same component are not allowed.
  Here, we consider a simple model that generates, from a given
  ensemble of dependency structures, a statistical ensemble of sets of
  components, allowing for components to appear with any
  multiplicity. 
  Our model is a minimal extension that is memoryless,
  and therefore accessible to analytical calculations.
  A mean-field analytical approach (analogous to the ``Zipfian
  ensemble'' in the linguistics literature) captures the relevant laws
  describing the component statistics as we show by comparison with
  numerical computations.
  In particular, we recover a power-law Zipf rank plot, with a set of
  core components, and a Heaps law displaying three consecutive
  regimes (linear, sub-linear and saturating) that we characterize
  quantitatively.
\end{abstract}

\maketitle

\section{Introduction}

Several physical, biological, and artificial systems are modular
and can be partitioned in well-defined basic components, revealing
essential features of their architecture. For example, genomes can be
regarded as sets of genes, operating systems as sets of packages,
texts as sets of
words~\cite{vanNim03,Gherardi:2013PNAS,Pang2013,Altmann2016,Mazzolini2017}.
A system showing this simple modular structure can be considered as a
specific collection of partitions of elementary components (a
``component system''), whose statistical properties may reveal
information on the system's generative mechanisms.  For example, books
in a linguistic corpus
composed by $N$ words (the total number of ``tokens'' in
quantitative-linguistics language) are
partitions of $N$ elements in component classes identified by the
different words, or ``types''.  These partitions, and their
statistical properties, are generally an extension of the classic
partitions intensively studied in probability~\cite{Pitman} as well as
in statistical mechanics, for example related to the equilibrium
statistics of particles in energy states~\cite{Leinaas1977}, or to
non-equilibrium site occupancy in driven diffusive
systems~\cite{Evans2005}, or resulting from stochastic processes based
on duplication-innovation mechanisms~\cite{angelini,Cosentino09}.

The representation of complex systems as component systems reveals a
variety of quantitative laws, which, intriguingly, are often conserved
across very different systems.  Prominent examples are Zipf's law,
concerning the power-law distribution of component
frequencies~\cite{zipf1935psycho,
  newman2005power,mitzenmacher2004brief, lotka1926frequency, Huynen98,
  axtell2001zipf}, and the sublinear scaling of the number of
different component classes with system size, often referred to as
Heaps' law
\cite{Herdan1960,heaps1978information,Powers1998,Cosentino09}.
The analysis of these laws has a long tradition in quantitative
linguistics~\cite{Altmann2016,zipf1935psycho}.  Several mechanisms of
text generation, based on different hypotheses on the fundamental
structures or principles of natural language, have been proposed to
explain these statistical
patterns~\cite{chancho2003,zanette2005dynamics,
  gerlach2013stochastic}.  Analogously, in genomics, the
coarse-grained view of genomes as component systems reveals emergent
quantitative invariants pointing to relevant underlying evolutionary
and architectural properties of
genomes~\cite{Koonin2011,Grilli2012,Lobkovsky2013,Maslov2009,Pang2013,vanNim03}.

Scale invariance (and universality), suggested by the presence of
Zipf's law, may be a natural consequence of
criticality~\cite{Mora2011}, either due to evolutionary tuning or
self-organization~\cite{bak1987self,Hidalgo2014}, as is well understood
via the renormalization group in statistical mechanics.  However, more
and more often scale-free features in different contexts have been
recognized as a possible consequence of stochastic processes based on
specific features, such as preferential
attachment~\cite{newman2005power,mitzenmacher2004brief} or a
history-dependent entropic reduction of the accessible
states~\cite{Hanel2011,Corominas-Murtra2015}.  The plethora of
possible mechanisms that can generate these pervasive statistical
patterns raises questions concerning their true origin and their
robustness~\cite{Sornette2006}.

Typically, these proposed mechanisms do not explicitly take into
account the functional specificities of single components,
nor their functional dependencies, nor the possible synergy and conflicts between
components. 
These interactions are essential in most empirical systems.
For example, the importance of this aspect is clear in operating
systems (or other large software projects), where a package performs a
specific function and typically requires the presence of other
components for functioning.  Indeed, models based on dependency
structures have emerged recently as a promising framework to
rationalize component dependencies and analyze their consequences in
terms of component statistics~\cite{Pang2013}.
Similar dependency structures also emerge in preference
prediction~\cite{Heckerman2000}, or for addressing causality in
financial data~\cite{Kenett2010}.

More precisely, a dependency structure is a directed graph (most
often, but not necessarily, acyclic), whose nodes are the components
(e.g., Linux packages, or genes) and whose links are the dependency
relations occurring between them (e.g. requirement constraints or
regulatory pathways).  A component depends on another if it is not
functional unless the latter is present.
A simple mechanism to build a component system compatible with a given
dependency structure identifies a system realization (e.g., a genome
or a book) with the choice of a node and all its direct and indirect
dependencies~\cite{Pang2013}.  This model allows to link quantitative
laws of the component statistics to topological properties of the
dependency structure (and hence to the generative processes sculpting
it).  For instance, a broad ensemble of dependency structures has the
property that the number of total dependencies of each node is scale
free in the limit of large system size (notice that this is a weaker
condition than the power-law distribution of degrees, i.e., of only
direct dependencies).

This topological property explains the fat-tailed distribution of
component occurrences across realizations, both in genomes and
operating systems~\cite{Pang2013}.
However, since this simple model was inspired by software packages, it
was developed in the restrictive case of binary (presence/absence)
abundances of components.  While a piece of software is either installed or
absent, in most empirical component systems a component type can be
present in many instances. For example, a gene is typically present in
multiple copies (called paralogs) in a genome, and the same word is
typically used several times in a single book.

Here, to capture this large class of component systems, we consider an
extension of the model proposed in Ref.~\cite{Pang2013} to the cases
in which components can appear with arbitrary abundances.  This
extension is able to address the question of how dependency structures
affect abundance-related features, such as Zipf's laws and in
particular Heaps' law. The same question is not defined in the case of
binary component abundances, where, for example, Heaps' law is
trivially linear.

\section{Model}

We consider a simplified description of complex functional
architectures as unordered sets of modular components.
Let $\mathcal U$ be the set of all unique components (the
\emph{universe}), and let $U=\left|\mathcal U\right|$ denote its
cardinality.  A \emph{realization} of this ``component system'' is a
set $r=\left\{c_i\right\}$ of components $c_i\in\mathcal U$, with
$i=0,\ldots,N$, where $N$ is the \emph{size} of the realization.
The rules for constructing a realization are specified by a network of
dependencies, as we explain below.

\begin{figure}[t]
\centering
\includegraphics[scale=1.1]{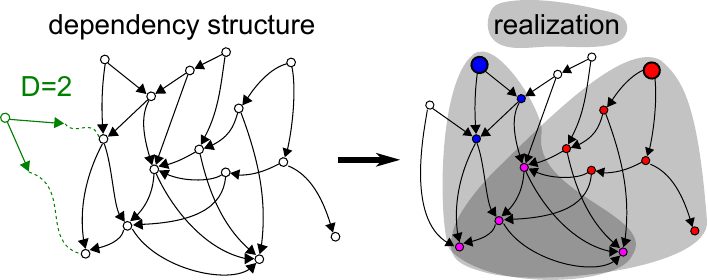}
\caption{\textbf{Illustration of the model. } The model is defined by
  two steps.  The first step creates a dependency structure by an
  incremental node-addition process with mean out-degree D.  The
  second step builds realizations drawing at random $k$ precursor
  nodes (larger circles in figure, where $k=2$) and taking all
  components belonging to their forward ``cones'' of dependency
  (shaded in gray).  This second step assigns multiplicities to the
  components appearing in each realization: the abundance of component
  $i$ is the number of precursors whose forward cones contain
  $i$.  In the realization illustrated in the right panel, two
  precursor components (larger circles) are chosen; the nodes
  (components) in the light-gray regions have multiplicity 1, while
  those in the dark-gray region have multiplicity 2.  }
\label{figure:model}
\end{figure}

A \emph{dependency structure} is a directed acyclic graph $\mathcal G$
on $\mathcal U$, which encodes the dependencies between the
components.
An edge $i\to j$ between two nodes $i$ and $j$ represents the relation
``$i$ depends on $j$''.  A component $i$ is said to depend on another
component $j$ if $i$ is not functional without $j$.  In empirical
cases, such a relation can be more or less strict depending on the
system; for instance it is enforced in software operating systems,
where a package cannot function unless all its dependencies are
installed, but not in metabolic networks, where alternative pathways
can be followed to the same metabolite~\cite{Fortuna2011,Maslov2009}.
The model assumes strict unbroken dependencies.
Notice that acyclicity of $\mathcal G$ is not stringently necessary;
however, as will be clear in the following, a cycle in $\mathcal G$
would behave as a single node in the model.

The topology of the dependency structure is conceptually separated
from the procedure that generates realizations satisfying the
dependency constraints. Here we use the ensemble of dependency
structures introduced in Ref.~\cite{Pang2013}, and we define a novel
method to build the realizations.
Specifically, as sketched in Fig.~\ref{figure:model}, the growth
process that creates the dependency network is an incremental
node-addition process generating structures with power-law distributed
sizes of direct and indirect dependencies (such property is crucial to
reproduce Zipf's law, see below).
More specifically, let us fix an average out-degree $D\geq 1$, i.e., an
average number of direct dependencies of a given component.  Starting
with an initial graph consisting of a single node, the full graph is
built node by node, by attaching the new node to $d+1$ randomly chosen
existing nodes (possibly with repetitions), where $d$ is a Poissonian
random variable of mean $D-1$.  The process is stopped when the
network reaches the predetermined size $U$.
A graph assembled with these rules is directed and acyclic, and hence a
good dependency structure, as can be seen by labeling each node by
the time $t=1,\ldots,U$ it was added to the network, and noticing that
there can be no links $t\to t'$ with $t<t'$.
Given a node $c$, the set $\wedge(c)\subseteq\mathcal{U}$ is defined
as the set of all nodes $c'$ such that there exists at least one
directed path in $\mathcal{G}$ starting from $c$ and arriving at $c'$.
We will call the set $\wedge(c)$ the \emph{forward cone} of $c$.
Similarly, we define the \emph{backward cone} $\vee(c)$ of $c$ as the
set of all nodes $c'$ such that there exists a path from $c'$ to $c$.
In other words, $\wedge(c)$ is the set of all components required be
the (direct and indirect) dependencies of $c$, whereas $\vee(c)$ is
the set of the nodes that depend (directly or indirectly) on $c$.

Once a dependency structure $\mathcal G$ is established, realizations
of the model, i.e., sets of components, are generated by the following
procedure (see also Fig.~\ref{figure:model}).
Let us fix a positive integer $k$, which represents the number of
``precursor'' components determining a realization.  The $k$
precursors $\left\{p_j\right\}$, $j=1,\ldots,k$, are chosen randomly
and independently among the nodes of $\mathcal G$.  Then the
corresponding realization is produced by taking all components
belonging to the forward cones of the precursors.
To complete the model specification one needs a rule to choose the
multiplicities of the components.  We allow a component belonging to
multiple cones to appear in multiple copies.
Let us imagine that precursors are added one at a time to the
realization.  At the $j$-th step the existing realization $r_{j-1}$
(possibly empty, when $j=1$) is extended, by the addition of elements
from the cone $\wedge(p_j)$ of the precursor $p_j$:
\begin{equation}
r_j = r_{j-1} \cup \Delta_j, \quad \Delta_j \subseteq \wedge(p_j).
\end{equation}
The choice of the incrementing set $\Delta_j$ must be done so as to
satisfy the dependency relations, i.e., $r_j \supseteq \wedge(p_j)$.
Doing this at every step ensures that the final realization $r_k$ will
not have any unsatisfied dependency.
Other than that, $\Delta_j$ is in principle unconstrained, and it may
be a random variable even at fixed $p_j$ and $r_{j-1}$.
Here we make the simplest choice
\begin{equation}
\Delta_j=\wedge(p_j).
\end{equation}
This makes the process Markovian, in the sense that
$r_j \setminus r_{j-1}\equiv \Delta_j$ is independent of $r_{j-1}$.
The case $k=1$, when a realization is specified by a single precursor,
reproduces the model of~\cite{Pang2013}.

\section{Results}

Our description separates the relational constraints existing between
the components, such as dependency and incompatibility, from their
functional correlations, such as synergy, co-occurrence,
interchangeability, conflict, and so on.
For what concerns the functional correlations, our model is the
simplest null model where no correlations between components are
dictated, other than those arising from the dependencies.
An advantage of this model is its analytical tractability.  The form
of Zipf's law is basically the same as in the binary model, and the
mean-field analysis parallels that in \cite{Pang2013}.  The main
additional output of our extension is a non-trivial Heaps' law, which
is derived analytically below.

\subsection{Zipf's law and the occurrence-abundance relation}

We now set out to compute the distribution of component abundance emerging
from a set of realizations in this model, equivalent to the empirical
Zipf's law measured from a set of texts.

Given a set of $R$ realizations of a component system, the
``popularity'' of a component $i$ can be measured in two ways: by its
\emph{abundance} $a_i$ and by its \emph{occurrence} $o_i$.  The
abundance counts the number of times that $i$ appears in all
realizations (with multiplicities):
\begin{equation}\label{eq:def_abb}
a_i = \frac{1}{k R} \sum_{r} \sum_{c\in r} \delta_{c,i}.
\end{equation}
In the model, the maximum abundance of a component $i$ corresponds to
drawing $i$ each time a cone is selected, for each realization.  In
such a case, the double sum in (\ref{eq:def_abb}) is $kR$.  Therefore,
the abundance $a_i$ is normalized so that $0\leq a_i \leq 1$ (we will
call $a_i$ the \emph{relative} abundance when it is important to
stress that it is an intensive quantity).
The occurrence $o_i$ measures the fraction of realizations containing
the component, regardless of its abundance:
\begin{equation}\label{eq:def_occ}
o_i = \frac{1}{R} \sum_{r} \left[1-\prod_{c\in r} \left(1-\delta_{c,i}\right) \right].
\end{equation}
With this definition, the occurrence is normalized so that
$0\leq o_i \leq 1$.

Zipf's law is a statement about the rank-frequency relation of
components.  Specifically, the frequency of a component decreases as a
power of its rank $r$, $f_r \propto r^{-\gamma}$ (where the rank is $1$
for the most frequent component, $2$ for the second most frequent, and
so on) \cite{Altmann2016}.  The Zipf relation is expected to be
independent of the number of cones $k$ (at least for large systems).
In fact, the abundance $a_i$ of a given component in $R$ realizations
constructed with $k$ cones each has the same distribution as that in
$kR$ single-cone realizations, since the choices of the cones are
independent.  $a_i$ can be estimated as the probability of choosing a
cone that contains $i$, which is proportional to the size
$\left|\vee(i)\right|$ of the backward cone of $i$:
\begin{equation}
\label{eq:abundance_vee}
a_i = \frac{\left|\vee(i)\right|}{U}.
\end{equation}

\begin{figure}[t]
\centering
\includegraphics[scale=1.3]{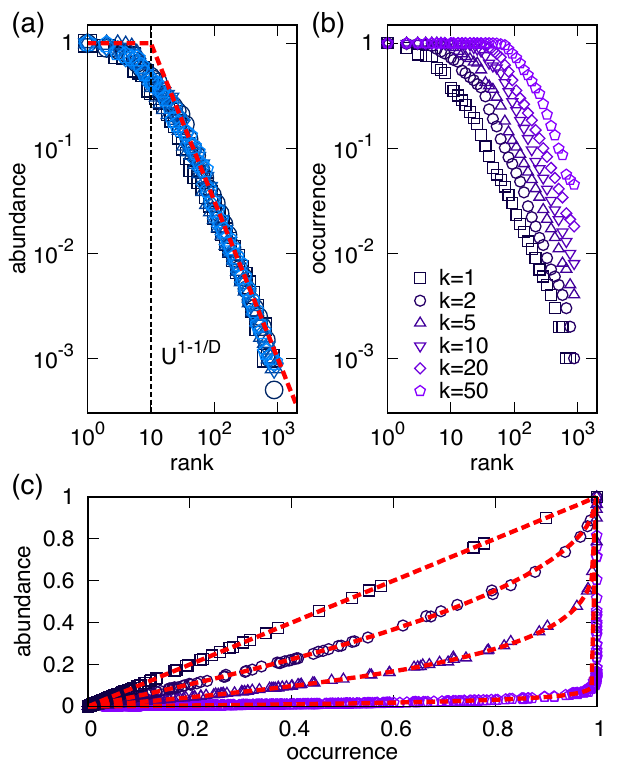}
\caption{\textbf{Simulations match the analytical prediction for the
    distribution of total abundance in a set of realizations
    (empirical Zipf's law) and for the abundance-occurrence relation.}
  (a) Rank-plot of component abundances.  (b) Rank-plot of component
  occurrences.  (c) Scatterplot of the abundance (y axis) versus the
  occurrence (x-axis) of each component.  Colored symbols are results
  of simulations ($1000$ realizations) with varying number of
  precursors $k$, the dashed lines are the analytical prediction.
  (Parameters: $U=1000$, $D=1.5$.)}
\label{figure:zipf}
\end{figure}

Let us call $\mathrm{rank}(i)$ the rank of component $i$ when
all components are ranked by their abundance.
Following~\cite{Pang2013}, an approximate relation can be derived between
$\left|\vee(i)\right|$ and $\mathrm{rank}(i)$,
which will allow to obtain an analytical estimate of Zipf's plot.
The $t$-th node in the network
(the one added at the $t$-th step of the construction,
when a network of size $t-1$ has been already generated)
has approximately $(U/t)^D$ nodes that depend on it.
This result can be obtained by writing an equation stating
that the backward cone of the $t$-th node is the union
of the backward cones of all the nodes that, at later times $t'$,
will directly attach to the $t$-th node.
Neglecting the intersections between these cones allows one to write
the recursion
\begin{equation}
\label{eq:cones_recursion}
\left|\vee(t)\right| = 1 + \sum_{t'=t+1}^{U} \frac{D}{t'} \left|\vee(t')\right|,
\end{equation}
where the factor $D/t'$ estimates the probability that the $t'$-th
node attaches to the $t$-th node
(with a slight abuse of notation,
we write $\wedge(t)$ and $\vee(t)$ for the forward and backward cones of the $t$-th node).
By approximating the sum by an integral and taking a derivative with respect
to $t$, one obtains a differential equation that is solved by 
$\left|\vee(t)\right| = (U/t)^D$.
For small $t$, however, $(U/t)^D$ is greater than the size of the network $U$.
In fact, the relation can hold only down to a cutoff $t_\mathrm{min}$,
which can be estimated by the condition
that the whole network depends on the $t_\mathrm{min}$-th node, i.e.,
$(U/t_\mathrm{min})^D=U$, which gives $t_\mathrm{min}=U^{1-1/D}$.
For any node below $t_\mathrm{min}$, the size of its backward cone is $\approx U$:
\begin{equation}
\label{eq:backwardconezipf}
\left|\vee(t)\right| \approx
\begin{dcases}
U\quad & t<U^{1-1/D}
\\
\left(U/t\right)^D & t\geq U^{1-1/D}
\end{dcases}
\end{equation}
Equations (\ref{eq:abundance_vee}) and (\ref{eq:backwardconezipf})
imply that if node $i$ is the $t$-th node in the network growth process,
then $t=\mathrm{rank}(i)$.
(This identification does not hold for the first $U^{1-1/D}$ components,
but this does not influence the result since the size of their backward
cones are equal in this approximation.)
Therefore, one obtains
\begin{equation}
\label{eq:zipf}
a_i \approx
\begin{dcases}
1\quad & \mathrm{rank}(i)<U^{1-1/D}
\\
\mathrm{rank}(i)^{-D} U^{D-1} & \mathrm{rank}(i)\geq U^{1-1/D}.
\end{dcases}
\end{equation}
This relation has the form of a Zipf power-law (with exponent $-D$)
with an initial ``core set'' consisting of $U^{1-1/D}$ components
having similar abundances. 
Figure \ref{figure:zipf}(a) compares the analytical form (\ref{eq:zipf})
with the results of simulations, showing good accord, especially
in the behavior of the fat tail.
The transition between the core and the tail, instead,
is less sharp than predicted. This is due to the fact that the
relation $\left|\vee(t)\right|=(U/t)^D$ starts to break down
before reaching $U$, and saturates more smoothly than
in the approximation made above.
Importantly, the relation between rank and relative abundance
does not depend on the number of cones $k$,
in accord with the foregoing prediction.

Contrary to the binary case discussed in Ref.~\cite{Pang2013},
where abundance and occurrence are the same quantity
[as shown by Eq.~(\ref{eq:abundance_occurrence}) below],
the rank-occurrence and rank-abundance relations in our model
are different in general.
Figure~\ref{figure:zipf}(b) shows that the Zipfian plot for the occurrence
has a similar appearance to the usual one, but with a larger and larger
core as $k$ increases.
When plotted as a probability distribution function in linear scale,
the rank-occurrence relation has a characteristic U shape \cite{koonin2008genomics,
touchon2009organised}, that highlights a subset constituted
by a large number of rare components and a subset of
components shared by a large fraction of realizations,
usually referred to as the ``cloud'' and ``core'' sets respectively
\cite{Lobkovsky2013}.
How these features are related to Zipf's law
is explored in detail in  Ref.~\cite{Mazzolini2017}.

The occurrence-abundance relation predicted by our model turns out to
be universal (or ``null''), meaning that it is insensitive to the
explicit form of Zipf's law, to the detailed structure of the network,
and even to the size $U$ of the component universe.  In fact, we show
here that a simple probabilistic argument gives a relation that is
consistent with simulations of the full model.  In the limit of large
$R$, we can assume that the occurrence of a component $i$ is equal to
the probability of choosing $i$ at least once in a single realization:
$o_i=1-(1-a_i)^k$, hence
\begin{equation}\label{eq:abundance_occurrence}
a_i = 1-\left(1-o_i\right)^{1/k}.
\end{equation}
For realizations with a single precursor ($k=1$), abundance and
occurrence are equal.  As $k$ increases, more and more components
(with larger and larger occurrence) assume small abundances.  In the
large-$k$ limit, all components have zero (relative) abundance, except
those with occurrence 1.  Figure~\ref{figure:zipf}(c) shows that a
scatterplot of abundance versus occurrence in simulations perfectly
matches the theoretical prediction (\ref{eq:abundance_occurrence}).
The figure shows results for a single choice of $D$ and $U$, but we
checked that these parameters have no effect on the curves.

\subsection{Analytical form of Heaps' law}

We now ask about the change of the typical number of distinct
components with realization size.  The calculation of the number
$F(N)$ of unique components in a realization of size $N$ can be
performed in a mean-field approximation, where the correlations
between nodes are neglected.
We consider a process where a realization is generated by extracting
$N$ nodes independently.
The full model adds entire dependency cones at once.  However, one
expects that the size $N$ of a realization scales linearly with the
number of cones $k$, and we have verified with simulations that this
is indeed the case.
The assumption of independent extractions makes our mean-field
approach to evaluate Heaps' law analogous to the so-called ``Zipfian
ensemble'' of quantitative linguistics, in which realizations are
generated through random extractions of components with probabilities
proportional to their frequencies~\cite{Mazzolini2017,van2005formal,
  Font-Clos2015}, or through a Poisson process in which component
frequencies establish their arrival rates~\cite{eliazar2011growth}.
In this framework, Heaps' law is the natural result of the
heterogeneity in component frequencies described by Zipf's law.

The probability
\begin{equation}
p(t) = \frac{1}{\Omega} \left|\vee(t)\right|
\end{equation}
of drawing the node $t$ is
proportional to the size $\left|\vee(t)\right|$ of the node's backward cone.
In a continuous approximation, the normalization $\Omega$ can be fixed
by the condition $\int_0^\infty p(t)\;\mathrm{d}t = 1$, which yields
\begin{equation}
\label{eq:normalization}
\Omega =
U \frac{D U^{1-1/D}-1}{D-1}.
\end{equation}
Note that $\Omega>U$ whenever $U>1$ and $D>1$.

Let $p_1(t, n)$ be the probability that the $t$-th node in the network
is drawn for the first time when the system being constructed has size
$n$:
\begin{equation}
p_1(t,n) = p(t) \left[1-p(t)\right]^{n-1}.
\end{equation}
A mean-field estimate of $F$ can then be obtained as
\begin{equation}
F(N) = \sum_{n=1}^N \sum_{t=1}^U p_1(t,n) 
\approx \int_0^U \!\!\mathrm{d}t \sum_{n=1}^N p_1(t,n)
\end{equation}
The geometric sum in $n$ gives simply the probability $1-[1-p(t)]^N$
that the $t$-th node has been drawn at least once after $N$ steps.
The mean-field expression for Heaps' law is then given by the
following integral:
\begin{equation}
\label{eq:heapsintegral}
\begin{split}
F(N) &= \int_0^U\!\!\mathrm{d}t \left\{ 1-\left[ 1-p(t) \right]^N \right\}\\
&= U - U^{1-1/D} \left( 1-\frac{U}{\Omega} \right)^N - \mathcal{I}(N),
\end{split}
\end{equation}
where the first term ($U$) comes from the integral of $1$, and the
second and third terms are the contributions of the two regions in
(\ref{eq:backwardconezipf}).  The remaining integral
\begin{equation}
\label{eq:defIntegralI}
\mathcal{I}(N)=
\int_{U^{1-1/D}}^U\!\!\mathrm{d}t \left[ 1- \frac{1}{\Omega} \left( \frac{U}{t} \right)^D \right]^N
\end{equation}
can be evaluated with the change of variables $z = (U/t)^D/\Omega$,
which gives
\begin{equation}
\label{eq:zIntegralI}
\mathcal{I}(N)=
\frac{U}{D} \Omega^{-1/D} \int_{1/\Omega}^{U/\Omega} (1-z)^N z^{-1-1/D} \mathrm{d}z.
\end{equation}
By remembering that the primitive of
$(1-z)^\alpha z^\beta$ is $z^{\beta+1} \;_2F_1(-\alpha, \beta+1, \beta+2, z) /(\beta+1)$,
where $_2F_1$ is the Gauss hypergeometric function,
one finally obtains
\begin{equation}
\label{eq:heapshyper}
\begin{split}
F(N) &= U - U^{1-1/D} \left( 1-\frac{U}{\Omega} \right)^N\\
&\phantom{=} - \;_2F_1 \left( -N, -\frac{1}{D}, 1-\frac{1}{D}, \frac{1}{\Omega} \right) U\\
&\phantom{=} + \;_2F_1 \left( -N, -\frac{1}{D}, 1-\frac{1}{D}, \frac{U}{\Omega} \right) U^{1-1/D}.
\end{split}
\end{equation}
The hypergeometric function $_2F_1(-N,\cdot,\cdot,\cdot)$ is real when
$N$ is an integer.
Fig.~\ref{figure:heaps} shows that the analytical mean-field
expression (\ref{eq:heapshyper}) nicely matches the results of
numerical simulations of the model.

\subsection{Linear, sublinear, and saturation regimes of Heaps' law}
\label{section:regimes}

\begin{figure}[t]
\centering
\includegraphics[scale=1.3]{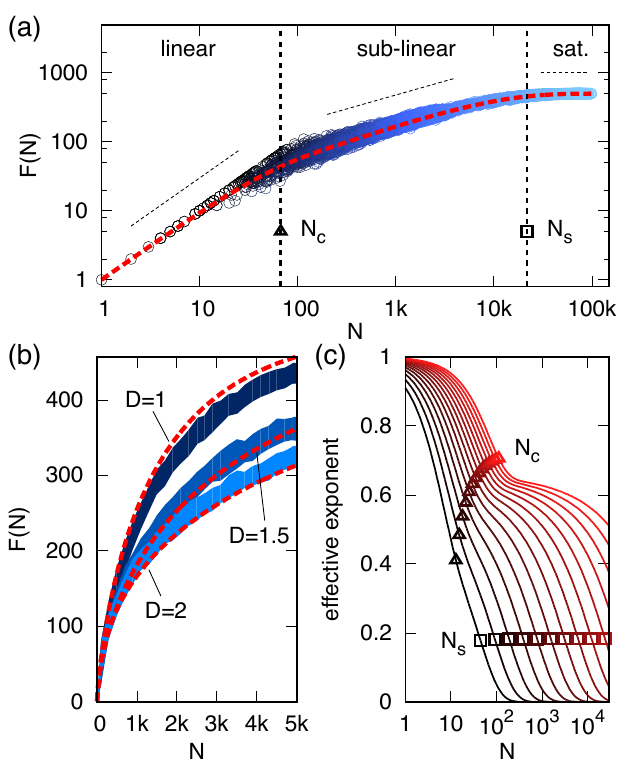}
\caption{\textbf{The model reproduces the characteristic three-regimes
  structure of Heaps' law.}  (a) The number of distinct components
  $F(N)$ of a realization as a function of realization size $N$ (in
  log-log scale).  The colored circles are simulations, lighter shades
  of blue correspond to larger numbers $k$ of precursors ($k$ ranges
  from $1$ to $3000$). The red dashed lines are the analytical
  prediction (\ref{eq:heapshyper}). The area of the solid curves
  represents the $90\%$ variability interval. (b) The same plot of $F(N)$,
  in linear scale, for three values of the mean out-degree $D$: $1$
  (dark blue), $1.5$ (blue) and $2$ (light blue). (Parameters:
  $U=500$, $D=2$, $13000$ realizations.) (c) Effective exponent versus
  realization size for for increasing values of $U$ 
  (black- to red-colored lines, ranging logarithmically from $U=10$ to $U\approx 6000$). 
  Triangles correspond to the analytical estimate of
  $N_\mathrm{c}$, which captures well the transition to a plateau
  region. The analytical estimate of $N_\mathrm{s}$ (squares)
  corresponds to an approximately constant effective exponent.}
\label{figure:heaps} 
\end{figure}

If a realization is constructed by incremental addition of randomly
chosen components, one expects $F(N)$ to be approximately linear for
small $N$, as it is unlikely to draw the same component twice.
Intuitively, the probability to do so increases with $N$, up to a
point where approximately all components in the universe will have
been included, and $F(N)$ will saturate to $U$.
For instance, this is the behavior observed empirically for the species-area relationship in ecology~\cite{rosindell2007species,grilli2012spatial,o2017cross}.
This behavior becomes apparent by plotting $F(N)$ in log-log scale
[see Fig.~\ref{figure:heaps}(a)].  There emerge three distinct regimes:
a linear increase for small $N$, a saturation to $U$ for large $N$,
and an intermediate regime where the sub-linear increase of $F(N)$
appears to be well described by a power law.
Two transition points can be identified, $N_\mathrm{c}$ and $N_\mathrm{s}$,
respectively at the crossover between the linear and the sub-linear regimes,
and at the onset of saturation.
We collect here a few analytical estimates and observations.

It is clear from expression (\ref{eq:heapsintegral}) that,
since $p(t)>0$ for a finite universe,
\begin{equation}
\lim_{N\to\infty} F(N) = U.
\end{equation}
This is a consequence of the definition of the model, whereby $F(N)$
is monotonic by construction and $F(N)\leq U$.  However, this limit is
not apparent from the final formula (\ref{eq:heapshyper}).  What
happens is that the (essential)
singularities of the two hypergeometric functions cancel out in the
large-$N$ limit.
This makes it difficult to compute values of $F(N)$ numerically in
this regime (see below).

An estimate of the point $N_\mathrm{s}$ where the saturation regime
sets in can be obtained from (\ref{eq:heapsintegral}).  The term with
$(1-U/\Omega)^N$ is significantly different from zero when
$U/\Omega \lesssim 1/N$, i.e., when $N \lesssim \Omega/U$.  The
integral $\mathcal{I}(N)$, instead, can be evaluated for large $N$ in
a saddle point approximation.  The integrand
[Eq.~(\ref{eq:defIntegralI})] attains its minimum at $t=U$, where it
is equal to $(1-1/\Omega)^N$; hence, it is significantly different
from zero when $N \lesssim \Omega$.  Therefore, both $N$-dependent
terms in (\ref{eq:heapsintegral}) are negligible when
$N \gtrsim N_\mathrm{s}=\Omega$, where $\Omega$ is given by
(\ref{eq:normalization}).

The small-$N$ behavior at finite $U$ can be obtained in principle from
Eq.~(\ref{eq:heapsintegral}) as well, by expanding in $N$ before
performing the integral in $\mathcal{I}(N)$.  However, it is easier to
analyze the onset of sub-linearity in the large-$U$ regime, by
expanding Eq.~(\ref{eq:heapshyper}) in powers of $1/U^{1-1/D}$.  This
can be done by using the definition of the hypergeometric function:
\begin{equation}
\label{eq:defhyper}
_2F_1(a,b,c,z) = \sum _{k=0}^{\infty } \frac{(a)_k (b)_k}{(c)_k k!} z^k,
\end{equation}
where $(\alpha)_k = \alpha(\alpha+1)(\alpha+2)\cdots(\alpha+k-1)$ is
the Pochhammer symbol.  The two terms of the form $_2F_1(...)$ in
(\ref{eq:heapshyper}) can be expanded in powers of $z$;
Eq.~(\ref{eq:defhyper}) shows that the term of order $z^j$ in such an
expansion is a polynomial of order $j$ in $N$.  The same property
holds for the small-$z$ expansion of the first $N$-dependent term in
(\ref{eq:heapshyper}), of the form $(1-z)^N$.  It is then easy to see,
by keeping track of the analytical and non-analytical powers of $U$,
that in the limit $U\to\infty$ the only non-vanishing term is $N$, and
Heaps' law reduces to the identity
\begin{equation}
\lim_{U\to\infty}F(N)=N.
\end{equation}
A linear onset is expected for small $N$ even when $U$ is finite.
Performing explicitly the expansion to first order in $U^{-1+1/D}$
yields
\begin{equation}
\label{eq:Fexpansion}
F(N) \approx N - \frac{1}{2}N(N-1) \frac{2(D-1)^2}{D(2D-1)} U^{-1+1/D}.
\end{equation}
The crossover point $N_\mathrm{c}$ separating the linear and sub-linear regimes
can be estimated by the point where (\ref{eq:Fexpansion}) reaches its maximum:
\begin{equation}
\label{eq:Ncrossover}
N_\mathrm{c} = \frac{D(2D-1)}{2(D-1)^2} U^{1-1/D}.
\end{equation}
This expression is expected to become inaccurate when $D\approx 1$
(where in fact it diverges), because all terms of order $U^{-j+j/D}$
with $j>1$, which are neglected in (\ref{eq:Fexpansion}), approach
constants for $D\to 1$.

In order to identify more visually the transition points
from the data, one can plot the \emph{effective exponent}
\begin{equation}
\gamma_\mathrm{eff} (N) =
\frac{\mathrm{d} \log F(N)}{\mathrm{d} \log N}
\end{equation}
which is easily computed from numerical data
as a discrete derivative.
$\gamma_\mathrm{eff}(N)$ measures the apparent exponent
that is obtained by approximating the function $F(N)$
locally by a power-law $N^{\gamma_\mathrm{eff}}$.
Figure \ref{figure:heaps}(c) shows the effective exponent for a range of
values of $U$.
For small $U$, the regimes are somewhat intertwined, and no sharp
transitions appear.  For larger $U$, $\gamma_\mathrm{eff}$ shows three
plateaux, corresponding to $\gamma_\mathrm{eff} = 1$,
$\gamma_\mathrm{eff} = 0$, and an intermediate value
$\gamma_\mathrm{eff} = \gamma$.
An approximate value for $\gamma$ can be obtained by making the approximation
$(1-\epsilon)^N\approx \exp(-N\epsilon)$ in Eq.~(\ref{eq:defIntegralI}),
where $\epsilon=(U/t)^D/\Omega$.
The integral then becomes
\begin{equation}
\label{eq:gamma}
\mathcal{I}(N)\approx \left(\frac{N}{\Omega}\right)^{1/D}
\left. \Gamma\left(1-\frac{1}{D},x\right) 
\right|^{x=N/\Omega}_{x=NU/\Omega}.
\end{equation}
In the intermediate regime $N_\mathrm{c}\ll N\ll N_\mathcal{s}$
the upper incomplete gamma function $\Gamma$ only contributes
a constant in $N$. Therefore from the prefactor one readily obtains
$\gamma=1/D$.
This same exponent can be derived in the framework of the Zipfian
ensemble computations by a simple scaling argument by assuming a pure
power-law behavior of $F(N)$ \cite{van2005formal, eliazar2011growth,
  lu2010zipf}.

Figure \ref{figure:heaps}(c) also shows that the transition points
computed above, i.e., $N_\mathrm{c}$ given by (\ref{eq:Ncrossover}),
and
\begin{equation}
N_\mathrm{s} = U \frac{D U^{1-1/D}-1}{D-1},
\end{equation}
are reasonable estimates of the sizes where the two regime shifts
occur.
Surprisingly, the estimate of $N_\mathrm{s}$ turns out to correspond
to an approximately $U$-independent value of the effective exponent.

\subsection{Stretched exponential saturation as a phenomenological
  expression of Heaps' law}

As pointed out above, the asymptotically flat behavior of Heaps' law
$F(N)$ in this model results from the cancellation of two infinite
contributions in the analytical formula.  This subtlety makes it
numerically challenging to evaluate $F(N)$, especially for large $U$
and $N$.  Such a difficulty prevents the use of
Eq.~(\ref{eq:heapshyper}) to estimate the parameters by fitting
against empirical data.
However, the analytical expression (\ref{eq:zIntegralI}) suggests a
simple phenomenological expression, which can be useful in fits.
Since the integration variable $z$ is small for large $U$, one can
attempt to approximate the integrand in $\mathcal{I}(N)$ by
$z^{-1-1/D} \exp(-zN) \mathrm{d}z$.  In this form, the integral is
similar to a representation of the stretched exponential function
$\psi_{\gamma,a}(x)=\exp(-a x^\gamma)$ in terms of exponential decays
as a Laplace transform
\begin{equation}
\label{eq:stretched_integral_representation}
\psi_{\gamma,a}(x)=\int_0^\infty P_{\gamma,a}(z) e^{-z x}\mathrm{d}z \
.
\end{equation}
The asymptotic behavior of $P_{\gamma,a}(z)$ is known to be
\begin{equation}
\label{eq:large_z}
P_{\gamma,a}(z)\sim z^{\gamma+1}
\end{equation}
for large $z$, and an exponential decrease for small
$z$~\cite{Johnston2006}.  This suggests the following phenomenological
expression:
\begin{equation}
\label{eq:stretched_exp}
F_\mathrm{ph}(N)=U\left[ 1-\exp\left( -a N^\gamma \right) \right] \ .
\end{equation}

\begin{figure}[t]
\centering
\includegraphics[scale=1.3]{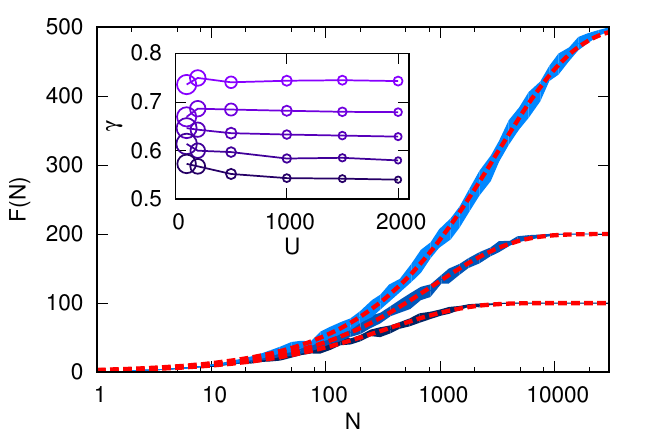}
\caption{\label{figure:stretched} \textbf{The stretched exponential is
    a good approximation of the Heaps plot generated by the model.}
  Number of unique components ($y$ axis, in linear scale) as a
  function of realization size ($x$ axis, in logarithmic scale), for
  three different values of $U$, the size of the universe.
  Simulations are the solid curves (dark blue $U=100$, blue $U=200$
  and light blue $U=500$; line width is the $90\%$ variability
  interval).  The dashed red lines are the analytical prediction.
  (Parameters: $D=1.5$, $k$ ranging from 1 to 5000.)  (Inset) The fitted
  stretched-exponential exponent $\gamma$ as a function of $U$.
  Different shades correspond to $D=1, 1.25, 1.5, 1.75, 2$, from top
  to bottom; point size represents errors.  }
\end{figure}

Figure \ref{figure:stretched} shows that the stretched-exponential
saturation, Eq.~(\ref{eq:stretched_exp}), is a remarkably good
approximation of the simulated data.  The log-linear scale reveals
that the agreement is tight on the whole range of $N$.  However, the
phenomenological expression fails to capture the transient linear
increase at small sizes (see Sec.~\ref{section:regimes}).  Indeed, the
small-$N$ behavior of $F_\mathrm{ph}$ is
$F_\mathrm{ph}(N)\sim UaN^\gamma$.
Interestingly, if one extracts $\gamma$ by matching the large-$z$
power-law scaling in Eq.~(\ref{eq:large_z}) with the factor
$z^{-1-1/D}$ in Eq.~(\ref{eq:zIntegralI}), one obtains $\gamma = 1/D$,
which is the same exponent that was derived from Eq.~(\ref{eq:gamma}).
Note however that the integration range in (\ref{eq:zIntegralI}) is
very different from the one in the integral representation
(\ref{eq:stretched_integral_representation}) of $\psi_{\gamma,a}$,
that is $(0,\infty)$.  As a consequence, the fitted exponents $\gamma$
can deviate from the simple scaling relation $\gamma = 1/D$.
Altogether, these considerations suggest that it is quite surprising
that the stretched exponential can be such a good approximation.

\section{Discussion}

It is interesting to compare the results of this positive model, where the
component statistics are purely determined by component dependencies,
with more null views of Heaps' and Zipf's laws.
Our analytical computations for the rank-abundance, rank-occurrence,
and Heaps' relations were all performed in an ensemble where
correlations between different components are neglected.  The
agreement between the results obtained in such a mean-field
approximation and the numerical simulations of the full model reflects
the fact that the main statistics considered are not sensitive to the
correlations.
In particular, the approximate stretched-exponential form of Heaps'
law is expected to hold not only for this particular positive model,
but for any model yielding a Zipf law similar to Eq.~(\ref{eq:zipf}).
This observation is in agreement with the results of
Ref.~\cite{Mazzolini2017}, where it is shown that building
realizations of a component system by randomly sampling from the
global frequency distribution of its components reproduces some
empirical statistics of real systems to a surprising detail.

\begin{figure}[t]
\centering
\includegraphics[scale=1.3]{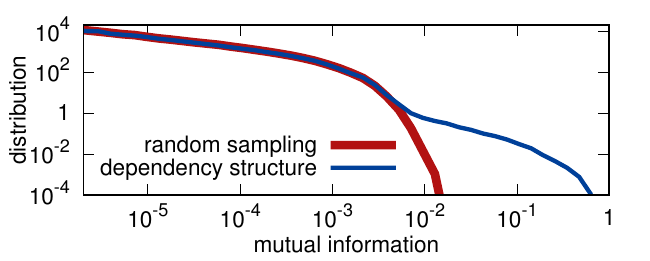}
\caption{\label{figure:mutual_info} \textbf{Non-null signatures of
    component dependency.} The distribution of mutual information ($x$
  axis), Eq.~(\ref{eq:mutual_info}), for the random sampling protocol
  of Ref.~\cite{Mazzolini2017} (thick red line) and for the model
  studied here (thin blue line). The frequency distribution of components
  was obtained by the positive model (with parameters $U=1000$,
  $D=1.5$, $k=50$), and $I(i,j)$ was computed for $R=1000$
  realizations of both models. The distribution ($y$ axis) is the
  normalized empirical distribution for all realizations and all pairs
  $(i,j)$.  }
\end{figure}

In this perspective, the positive framework established here around
the concept of dependency structures introduces positive features
without violating many of the null predictions of the random sampling.
Understanding how the positive trends emerge from the random features
is clearly crucial to determine the role played by dependencies in a
given empirical system. We briefly discuss two such positive features
here, leaving a more detailed investigation for future work.
First, the Zipf law in Eq.~(\ref{eq:zipf}), which is the main input
of a sampling procedure~\cite{Mazzolini2017}, is instead a positive
\emph{output} of the model studied here: fixing the network of
dependencies between components constrains all the main statistics
regarding their abundance and occurrence, including their frequency
distribution.
Second, the existence of dependency relations generates nontrivial
correlations between the components.  These correlations can be
observed, e.g., by measuring the empirical distribution of the mutual
information between the occurrences of pairs of components.
Let $p_i(x)$ be the fraction of realizations in which the component $i$
has state $x$, where $x=1$ means it is present and $x=0$ means it is
absent, and let $p_{ij}(x,y)$ be the fraction of realizations in which the
component $i$ has state $x$ and the component $j$ has state $y$.  The
mutual information between the two components, measuring how much the
presence of one informs on the presence of the other, is defined as
\begin{equation}
\label{eq:mutual_info}
I(i, j) = \sum_{x,y} p_{ij}(x,y) 
\log{\frac{p_{ij}(x,y)}{p_i(x) p_j(y)}}.
\end{equation}
If the probabilities are factorized, i.e., $p_{ij}(x,y)=p_i(x)p_j(y)$,
as it is the case in the random sampling model, then $I(i,j)=0$.  If
the empirical probabilities are computed from $R$ samples, one expects
$I(i,j)$ to converge to $0$ in the large-$R$ limit; yet it will have a
non-trivial distribution for finite $R$.
Figure~\ref{figure:mutual_info} shows the distribution of the mutual
information $I(i,j)$, averaged over all pairs $i$ and $j$, for a set
of realizations generated from random sampling, compared to the
positive model based on dependency structures studied here.  It is
clear that, beyond a common bulk due to finite-size fluctuations, the
two models have very different profiles in terms of pairwise
correlations.  The effect of the underlying dependency structure
generates a high mutual information tail, populated by highly correlated
pairs of components, and emerging from a background of pairwise
relationships compatible with the null random-sampling model.
We also expect that dependency structures will create correlations
that go beyond the pairwise relations between components.

The model presented here provides the simplest generative mechanism
producing collections of components consistent with a given dependency
structure.
Our construction extends the one proposed in Ref.~\cite{Pang2013} to
the case of components with non-binary abundance.
While the distribution of component abundances has a power-law tail
with an initial core, like in the case of Ref.~\cite{Pang2013}, the
situation is more complex here, as the distribution of component
abundances does not coincide with that of occurrences, due to the
non-binary nature of the multiplicities.
More general families of models can be defined by specifying the rules
for selecting the abundance of nodes belonging to the dependency cones
of more than one precursor.
The additive choice taken here, where each cone determines the
addition of one copy of each component belonging to it, provides a
minimal model that is memoryless, and therefore still accessible
analytically.

Since a wide variety of systems can be represented as collections of
components belonging to a common pool, our model has general
applicability.  A special case is that of genomes, which deserves some
observations.
A dependency structure between genes represents the recipes binding
the functional roles of different protein families, thereby
determining their usefulness in the same genome. For example, a gene
could depend on another if it is found downstream in the same
metabolic pathway~\cite{Pang2013}. The topology of such dependency has
not been fully characterized. Likely, it comprises both feedforward
and feedback structures, as well as non-directed exclusion principles
(whereby a gene might not be necessary or useful if another one is
present).
Concerning the additive choice discussed above, it is possible (and
likely) that the choice of a memoryless process is too restrictive in
the context of genomics.  For instance, not all gene families required
by the presence of multiple precursors need to be present in multiple
copies.
Future investigations could aim at defining more stringently from data
the minimal features of a model of dependency that could realistically
describe genomes. This could be inferred by the correlation structure
of domain abundances from sets of entirely sequenced genomes.


\acknowledgements
We thank 
Michele Caselle,
Amos Maritan,
and Sergei Maslov
for useful discussions and advice.

\end{document}